\documentclass[fleqn,usenatbib]{mnras}

\usepackage{newtxtext,newtxmath}

\usepackage[T1]{fontenc}

\DeclareRobustCommand{\VAN}[3]{#2}
\let\VANthebibliography\thebibliography
\def\thebibliography{\DeclareRobustCommand{\VAN}[3]{##3}\VANthebibliography}

\usepackage{graphicx}	
\usepackage{amsmath}

\hypersetup{pdfauthor={Name}}
\usepackage{amsfonts}

\title[Generation of Hyperbolic Trajectories]{A Two-Dimensional Analytic Solution for the Generation of Hyperbolic Trajectories Via A Single Close Encounter with Applications To Interstellar Objects}

\author[Monk \& Seligman]{
Hayden Monk,$^{1}$\thanks{E-mail: monkhayd@msu.edu}
Darryl Z. Seligman$^{1}$
\\
$^{1}$Dept. of Physics and Astronomy, Michigan State University, East Lansing, MI 48824, USA\\
}

\begin{document}
\label{firstpage}
\pagerange{\pageref{firstpage}--\pageref{lastpage}}
\maketitle

\begin{abstract}

The discovery of interstellar interlopers such as 1I/‘Oumuamua, 2I/Borisov, and 3I/ATLAS have highlighted the necessity  of understanding the dynamical pathways that eject small bodies from planetary systems into hyperbolic trajectories. In this paper we examine the orbital elements of particles in the restricted three-body problem prior to  and post  scattering onto hyperbolic trajectories by massive perturbers. Building on previous work, we calculate  closed-form --- but approximate --- analytic criteria that map pre- to post-encounter orbital elements. An application of these equations demonstrates that ejection occurs most efficiently when the orbital eccentricity of the massless test particle exceeds a minimum threshold, $e\gtrsim0.4$. The primary driver of the final eccentricity is the component of the perturber-centric velocity projected along the direction of motion of the perturber. These analytic criteria  are then benchmarked and validated against numerical simulations which demonstrate that they provide a reasonably good zeroth-order approximation for ejection behavior. However, system-specific cases will generally require numerical simulations in addition to this analytic construction. The methodology is applied to (i) the solar system and exoplanetary systems (ii) $\beta$ Pictoris and (iii) HR 8799 to evaluate the pre-scattering orbits of ejected particles. This method provides a transparent and computationally efficient tool for identifying orbits within a given system from which interstellar objects are efficiently ejected via a single scattering event from a massive perturber.
\end{abstract}


\section{Introduction}
There have been three $\sim$100-meter-scale interstellar objects discovered passing through the Solar System: 1I/'Oumuamua \citep{Williams17}, 2I/Borisov \citep{Borisov_2019_discovery}, and 3I/ATLAS \citep{Denneau2025}.  The discovery of 1I/'Oumuamua  specifically implies that the spatial number density of the population is $\sim0.1$ au$^{-3}$  \citep{Jewitt2017,Trilling2017,Laughlin2017,Do2018,Levine2021}. Displaying an unusually red spectrum and a 6:6:1 aspect ratio \citep{Mashchenko2019}, 1I/'Oumuamua also showed no sign of a cometary tail and was photometrically inactive \citep{Meech2017,Ye2017,Jewitt2017,Trilling2018}. Both 2I/Borisov and 3I/ATLAS exhibited a pronounced comet tail and clear coma \citep{Jewitt2019b,Fitzsimmons2019,Bolin2019,Ye2020,McKay2020,Guzik2020,Hui2020,Kim2020,Cremonese2020,yang2021, jewitt2025hubblespacetelescopeobservations}.  1I/'Oumuamua displayed comet-like nongravitational acceleration \citep{Micheli2018}. While radiation pressure is a feasible explanation \citep{MoroMartin2019, Flekkoy19,Luu20}, \citet{Micheli2018} attributed the nongravitational acceleration exhibited by 1I/'Oumuamua to outgassing. Upper limits on production of volatile species of 1I/'Oumuamua were limited, but carbon-based species and some H$_2$O daughter products were constrained \citep{Ye2017,Park2018,Trilling2018,Seligman2021}; see Table 3 in \citet{Jewitt2023ARAA}. Since its discovery, a population of photometrically inactive near-Earth objects (NEOs) with significant nongravitational acceleration has been observed \citep{Farnocchia2023,Seligman2023b,Seligman2024PNAS}.

3I/ATLAS is the third interstellar object and was discovered in July 2025. Early observations of 3I/ATLAS revealed distant cometary activity \citep{Jewitt2025, Alarcon2025, Chandler2025Rubin3IATLAS, Frinke2025} and  a reddened color \citep{seligman25, Opitom2025, Belyakov2025, Kareta2025, Tonry2025}. Multiple authors identified tentative periodic signals possibly due to nucleus rotation \citep{Santana-Ros2025, Marcos2025}. Subsequent spectroscopic observations suggested a CO$_2$-dominated coma containing  water-ice grains \citep{Cordiner2025JWST, Yang2025, Lisse2025SPHEREx3IATLAS} as well as H$_2$O, HCN, Ni, OSC, CO, CN and CH$_3$ OH\citep{Xing2025, Coulson2025, Manzano2025, Rahatgaonkar2025, Hoogendam2025,Roth2025}. Pre-discovery images of 3I/ATLAS showed activity at distances of at least $\sim$ 6 au \citep{Feinstein2025PrecoveryTESS3IATLAS, MartinezPalomera2025TESS3IATLAS, Ye2025, Chandler2025Rubin3IATLAS}. The object is likely 3-11 Gyr old, with unperturbed kinematics over at least the last 10 Myr \citep{Perez-Couto2025,Taylor2025}.

    Several interstellar object production mechanisms have been invoked to explain the contrasting properties of 1I/'Oumuamua and 2I/Borisov. A straightforward progenitor scenario is ejection via  scattering by a massive perturber such as a giant planet. This mechanism is known to operate with efficacy in the solar system \citep{Nesvorny2018}, as well as in high-multiplicity, compact systems \citep{Albrow2025}. \citet{Opik1976} provided an analytic framework for predicting the outcomes of such scattering events in the restricted three-body problem. The methodology allows for closed-form solutions by treating the interaction like a two-body interaction over a short period of time. It has been expanded with analytic and numerical methods \citep{Carusi1990, Wetherill_Cox_1984, Greenberg1988, Valsecchi1997, Valsecchi2003resonant, Valsecchi2018}.

    Examination of ejections in the three-body problem have been conducted in a number of related fields. One of the most closely related fields of study is that of the formation of the Oort cloud, since it largely focuses on scattering by large planets \citep{Wyatt2017_sweetspot, Duncan1987, Fernandez1997}. Particle ejections are also relevant for planet-planet interactions, the generation of free floating planets \citep{Raymond2009,Huang2025FreeFloatingMoons}, and star-star or star-black hole interactions that generate hypervelocity stars \cite{Yu2003, Wang2023, Baumgardt2006}. 
    
    In this paper we extend this methodology and apply it to identify planetesimals within a given system that are likely to be ejected based on their  orbit immediately prior to a scattering event. \citet{Huang2025} presented a related analysis in the context of repeated scattering events. In many cases, ejection occurs as the result of many close encounters, and so our method applies specifically to the event that causes the orbit to transition from elliptical to hyperbolic immediately prior to ejection.

\section{Methods}

In this paper we consider a three-body system consisting of a central star, a planet, and a massless particle. The massless particle has  orbital elements that we refer to as unprimed $a, e, i, \omega, \Omega$. Primed variables indicate the post-encounter parameters for the massless test particle for the remainder of this paper. These correspond to the semimajor axis, eccentricity, inclination, argument of pariapsis, and longitude of the ascending node, respectively. We assume the perturbing planet has a circular orbit, and the particle orbits on an eccentric, crossing orbit coplanar with the planet such that $i=0$. A crossing orbit necessitates $q<a_p<Q$, where \textit{q} and \textit{Q} are the perihelion and aphelion distance of the particle orbit and $a_P$ is the semimajor axis of the planet. Our coordinate system is defined such that $\omega=\Omega=0$. We calculate  how the orbit of the particle determines the outcome of a scattering event with the planet. To simplify analysis, we set $k\sqrt{M}=1$, where $k$ is the Gaussian gravitational constant.

\subsection{Pre-encounter Parameters}

The asymptotic, planet-centric velocity vector of the particle, $\vec{U}$ is  described by \citet{Carusi1990}, and also appears in Equation 8 in \citet{Huang2025}.  We use units of $a_P$, the semimajor axis of the planet. The x-axis is  oriented radially away from the star,  and the y-axis is directed along the instantaneous velocity vector of the planet\footnote{The z-axis is aligned with the angular momentum vector of the planet, although $U_z=0$ for the rest of this paper.}. The velocity is described as:

\begin{equation}\label{eq:U_component_matrix}
\begin{pmatrix}
U_x \\
U_y \\
\end{pmatrix}
\equiv
\begin{pmatrix}
\pm\sqrt{2-1/a-a(1-e^2)}\\
\sqrt{a(1-e^2)}-1\\
\end{pmatrix} \,.
\end{equation}

We can also define the direction of $\vec{U}$ in angular coordinates using $\phi$ and $\theta$, where $\phi\in[+\pi/2,-\pi/2]$ defines the azimuthal angle measured from the positive z-axis, and $\theta\in[0,\pi]$ defines the polar angle, measured from the positive y-axis within the ecliptic plane\footnote{Notably, this does not describe the entire sphere, but half a sphere is sufficient to examine the dynamics, which can be extended by symmetry.}. $\phi$ can take only the values 0 or $\pi$, describing the sign of the angular coordinate $\theta$.
\begin{equation}\label{eq:phi_theta_matrix}
\begin{pmatrix}
U_x \\
U_y \\
\end{pmatrix}
\equiv
\begin{pmatrix}
U\sin{\theta}\sin{\phi}\\
U\cos{\theta}\\
\end{pmatrix}\,.
\end{equation}

$\theta$ can be defined as a function of $a$ and $U$ (Equation 5 in \cite{Carusi1990}):

\begin{equation}\label{eq:cos_theta}
    \cos\theta \equiv \frac{1-U^2-1/a}{2U}\,.
\end{equation}

\subsection{Post-Encounter Parameters}

Once we have a robust description of the initial velocity vector, $\vec{U}$, we can directly generate the final velocity vector, $\vec{U}'$, along with the final orbital elements. By virtue of our focus on ejections, we are particularly interested in $e'$, the final eccentricity. We follow previous notation in which prime symbols denote a post-encounter variable. For example, the pre- and post-scattering semimajor axes are  $a$ and $a'$ respectively. The following equations  are based on \citet{Opik1951}. We calculate the kinematics of a perfect intersection of the orbits and apply them to a close encounter. The assumptions and range of applicability of this method are discussed in Section \ref{sec:assumptions} and the Introduction in \citet{Greenberg1988}. 

Our examination of $\vec{U}$ takes place within the planet-centric frame, where we treat the scattering as a two-body interaction, so the outgoing speed is equal to the incoming speed, i.e. $U=U'$. Because this removes a degree of freedom, we define $\vec{U}$ in angular coordinates and frame the scattering as a rotation from $\vec{U}$ to $\vec{U'}$, i.e. $(\theta,\phi)$ to $(\theta',\phi')$. The magnitude of this rotation is denoted by $\gamma$ and computed using the impact parameter of the interaction and the mass of the planet (Equation 10 in \cite{Carusi1990}):

\begin{equation}\label{eq:gamma_def}
\tan\,\big(\frac{\gamma}{2}\,\big)\equiv\frac{m}{bU^2}\,.
\end{equation}

In Equation \ref{eq:gamma_def}, $b$ is the impact parameter between the particle and planet in units of the planet's semimajor axis ($a_{P}$), and $m$ is the mass of the planet in solar masses. 
$\gamma$ is used to find $\theta'$, the post-encounter value of $\theta$, via the following equation (see Equation 11 of \cite{Carusi1990}):
\begin{equation}\label{eq:theta_prime_eq}
\cos\theta'\equiv\cos\theta\cos\gamma+ \sin\theta\sin\gamma\cos\psi\,.
\end{equation}

In Equation \ref{eq:theta_prime_eq}, $\psi$, the second angular descriptor of the three-dimensional transformation, is not neglected.  Note that the domain of $\theta'$ is $[0,\pi]$. $\psi$ describes the orientation of the deflection, while $\gamma$ describes the magnitude. $\psi$ is a free parameter because it represents the direction in which the particle passes by the planet \citep{Valsecchi2018, Valsecchi2003resonant}, and we approximate the solution as a perfect intersection of orbits. See descriptions in \citet{Carusi1990} and \citet{Valsecchi2005} for more information. Both $\psi=0$ and $\psi=\pi$ lie within the x-y plane. $\psi=0$ represents a rotation towards the y-axis in the x-y plane, while $\psi=\pi$ represents a rotation away from the y-axis. These correspond to trailing-side and leading-side scattering events, respectively. In the case of particles approaching from interior to the orbit of the planet, $\psi=0$ indicates a clockwise rotation, while $\psi=\pi$ indicates a counterclockwise rotation. The discrete nature of $\psi$ in two dimensions produces two simplified cases of Equation \ref{eq:theta_prime_eq}. We define $\cos\theta'^-=\cos(\gamma-\theta)$ for the $\psi=0$ case, and $\cos\theta'^+=\cos(\gamma+\theta)$ for the $\psi=\pi$ case.

The relationship between $\vec{U}$, $\vec{U}_y$, and $e$ is given by Equation 32 in \cite{Carusi1990}:
\begin{equation}\label{eq:e_prime_eq}
1-e^2=\big(1-U^2-2U_y\big)\,\big(1+U_y\big)^2\,.
\end{equation}

This relationship holds true both before and after the encounter, and therefore  also defines $e'$. $U$ and $\theta'$ define $U_y'$ and therefore the final eccentricity:

\begin{equation}\label{eq:e_prime_eq_solved}
    (e')^2=1-\bigg[\big(1-U^2-2U_y'\big)\,\big(1+U_y'\big)^2\bigg]\,.
\end{equation}

Note that final eccentricity does not depend on $\phi$, indicating that the sign of $U_x$ is irrelevant determining whether the particle is ejected. This symmetry about the y-axis is discussed in Section \ref{sec:jupiter_analog}.

\subsection{Direction of Scattering}
\label{subsec:branch_choice}

This section examines the relationship between leading- and trailing-side scatterings. $\theta'$ and $e'$ are related only though $U_y'$. To show that $e'$ is a monotonic function of $U_y'$, we differentiate Equation \ref{eq:e_prime_eq_solved} w.r.t.\ $U_y'$ to get:

\begin{equation}\label{eq:e_derivative}
\frac{\partial (e'^2)}{\partial U_y'} = 2\,\big(1 + U_y'\big)\,\big(U^2 + 3\,U_y'\big)\,.
\end{equation}

From Equation \ref{eq:e_prime_eq_solved}, $e'>1$ if and only if $U_y'>(1-U^2)/2$. Equation 3 in \cite{Carusi1990} establishes a relation between the magnitude of $\vec{U}$ and the Tisserand parameter:
\begin{equation}\label{eq:U_modulus}
    |U|\equiv\sqrt{3-T} \,.
\end{equation}
Also see Appendix A in \citet{Huang2025} for a more detailed derivation.
Because the Tisserand parameter is nonnegative for $i=0$, Equation \ref{eq:U_modulus} implies that $U\leq\sqrt{3}$, so $U_y'>-1$ for ejected particles. In this regime, the first term in Equation \ref{eq:e_derivative} is strictly nonnegative and the condition for $\frac{\partial (e'^2)}{\partial U_y'}>0$ is $U_y'>-U^2/3$. 
It is evident that $(1-U^2)/2\geq-U^2/3$ over the interval $[0,\sqrt{3}]$, so, for all values of $U$, an ejected particle will have $\frac{\partial (e'^2)}{\partial U_y'}>0$. The difference between the two cases of $U_y'$ is:

\begin{equation}\label{eq:Uy'_difference}
U_y'(\theta^-)-U_y'(\theta^+)=2\sin{\theta}\sin{\gamma}\,.
\end{equation}
$\theta$ and $\gamma$ are defined such that $\sin{\gamma}\geq0$ and $\sin{\theta}\geq0$, so $U_y'(\theta^-)\geq U_y'(\theta^+)$, and by the sign of the derivative, $e'(\theta'^-) \geq e'(\theta'^+)$ for ejected particles. Any particle ejected with $\psi=\pi$ will also be ejected with $\psi=0$, so we use the $\psi=0$ case to define $\theta'$. Geometrically, a value of $\psi=0$ is synonymous with the closest approach occurring along the $\zeta$-axis, which is oriented opposite to the projection of $\vec{U}$ onto the $b$-plane \citep{Valsecchi2003resonant}. The more energetically favorable interaction occurs when the closest approach of the particle is on the trailing side of the planet.  Equation \ref{eq:e_derivative} demonstrates that a trailing-side interaction will generally pull the particle into closer alignment with the velocity of the planet, increasing $U_y'$. 

\subsection{Discussion of Assumptions}\label{sec:assumptions}
Although this method introduces valuable simplifications to the three-body problem, it is important to understand its range of applicability. In this section we specify key assumptions in our methodology that define its scope. The scattering is assumed to be  a two-body interaction and therefore only applies exactly within the  Hill Sphere of the perturber.  \citet{Wetherill_Cox_1984} derived a minimum speed at which the particle must be traveling to render the methodology valid. Specifically, they claimed that the method was invalid when $V/V_e<0.35$, where $V$ is the relative velocity of the particle and $V_e$ is its escape velocity. \citet{Greenberg1988} disputed the use of this ratio as an assessment of validity, but agreed about the existence of a lower threshold. Second, this methodology only applies to the orbits of particles immediately before ejection \citep{Greenberg1988}. Finally, we assume that everything is coplanar and $i=0$ for all bodies. This assumption is discussed further in Section \ref{subsec:3d}. Our analysis presents a zeroth-order starting place to examine both the ejection efficiency of a system and the planetesimal reservoir to which the ejected particle belonged immediately prior to its ejection.

\begin{figure}
\centering
\includegraphics[width=1.\linewidth]{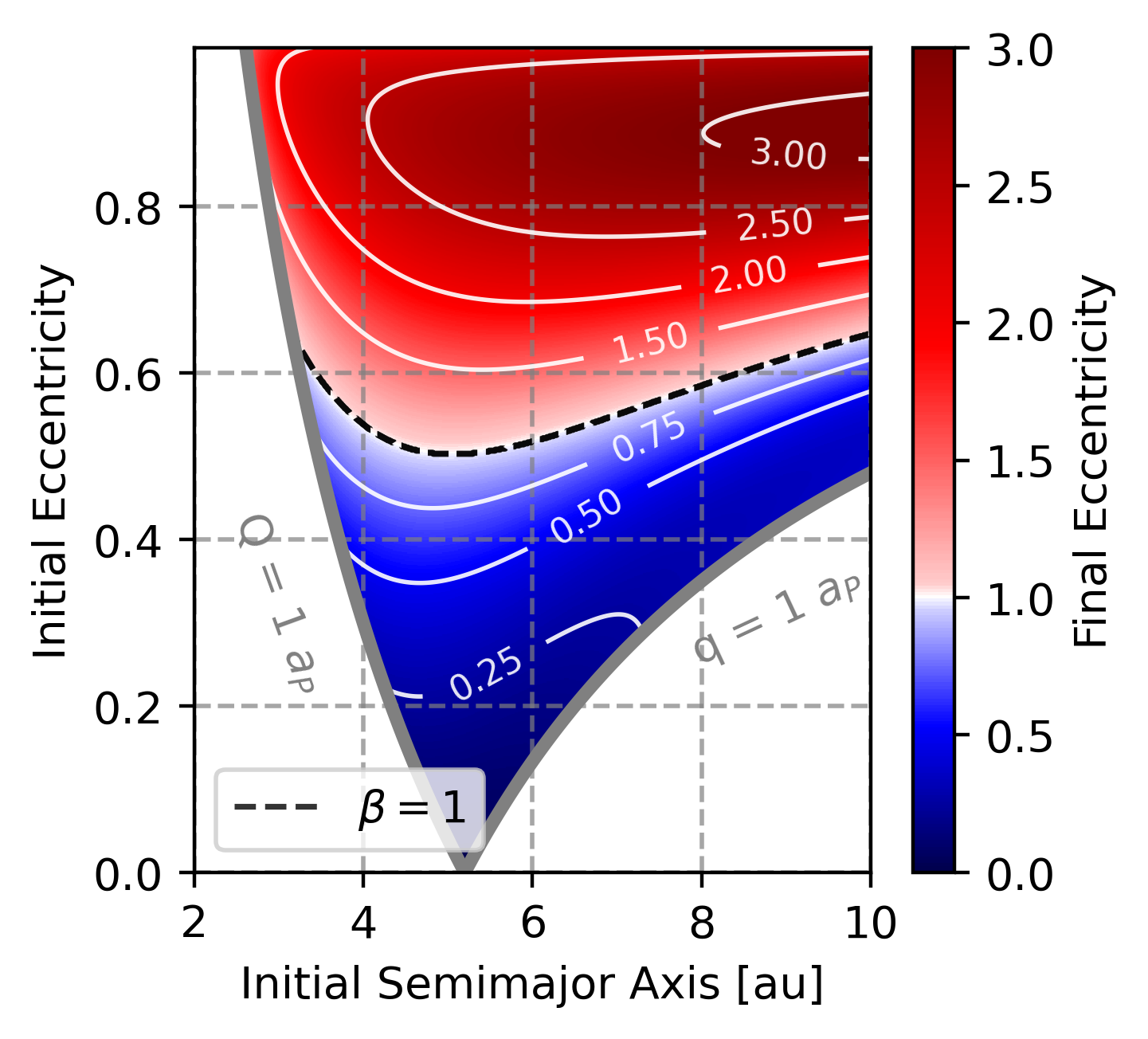}
\caption{Post-encounter eccentricity of a massless test particle that scatters off of a Jupiter analog. Color indicates  the final  eccentricity as a function of initial eccentricity and semimajor axis for interactions with a Jupiter-mass perturber at an impact parameter of 10 $R_J$. The blank sections of the parameter space correspond to non-crossing orbits, where $U_x$ is undefined by Equation \ref{eq:U_component_matrix}.}
\label{fig:jupiter_ejections}
\end{figure}
\begin{figure*}
\centering
\includegraphics[width=1.\linewidth]{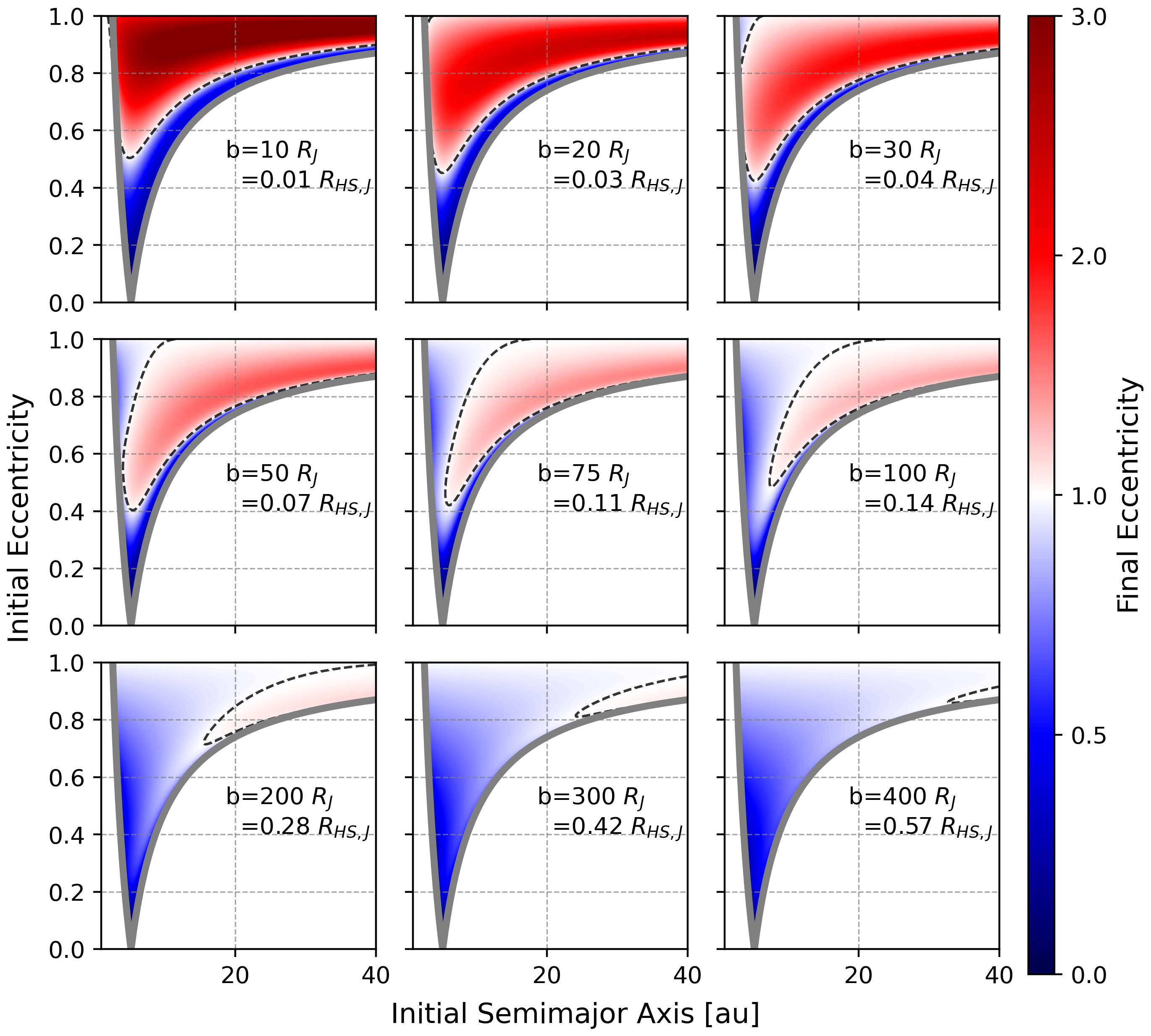}
\caption{Each subpanel is similar to that shown in Figure \ref{fig:jupiter_ejections} for a range of impact parameters between 10-400 $R_J$. The black, dashed lines follow $\beta=1$.  $e'$ decreases for more distant interactions. At 400 $R_J$, almost no orbits with $a\in[0, 40]$ au are ejected}
\label{fig:9panel}
\end{figure*}

\section{Ejecting Particles}\label{sec:ejected_particles}

\subsection{Ejection Condition}\label{subsec:ejection_condition}
From Equation \ref{eq:e_prime_eq}, it is evident that a given particle is ejected if 
\begin{equation}\label{eq:ej_condition}
1-U^2-2U'_y<0 \,.
\end{equation}
Therefore, a particle will be ejected following a close encounter with a massive perturber if: (a) it has a sufficiently large velocity and (b) the projection of its final velocity onto that of the perturber is sufficiently large. This agrees with the earlier analysis in Section \ref{subsec:branch_choice} that showed that an increase in the velocity's y-component correlates with an increase in eccentricity. Increased alignment with the y-axis represents a "kick" in energy in the star-centric frame. Beginning with $(1-(U')^2-2U'_y)<0$, we can rearrange to show its dependence on $\theta'$:
\begin{equation}\label{eq:e'_criterion}
    1-U^2-2U\cos{\theta'}<0\,.
\end{equation}

In terms of $\gamma$ and initial orbital properties, the necessary condition is:

\begin{equation}
1<U^2-2\,\big[U_y\cos(\gamma)+\big(U^2-U_y^2\big)^{1/2}\sin(\gamma)\,\big]\,.
\end{equation}

The expression in brackets is of the form $A\cos{\gamma}+B\sin{\gamma}$, which can be expressed in terms of amplitude and phase. We define the quantity $\beta$ as:

\begin{equation}\label{eq:beta_def}
    \beta=U^2+2U\cos(\gamma-\delta)\,.
\end{equation}

The phase shift $\delta$ is the angle between $\vec{U}$ and the negative y-axis, defined as

\begin{equation}\label{eq:delta_def}
    \tan{\delta}=-\, \bigg(\,\frac{|U_x|}{U_y} \, \bigg)\,\,.
\end{equation}

 $\beta$ is a dimensionless quantity which serves as a useful  criterion for ejection. If $\beta\ge1$, then $e'\ge1$. If $\beta<1$, then $e'<1$. This holds in three dimensions as well, as discussed in Section \ref{subsec:3d}.

\subsection{Fiducial Case: Jupiter Analog}\label{sec:jupiter_analog}

For a given perturber, $e'$ is a function of $a, e$, and $b$. Using a Jupiter analog planet with a fixed value of $b$, we calculate $e'$ in ($a,e$) space. The planet has $M=M_J$, $a=a_J$, and $e=0$. The subscript $J$ denotes that of Jupiter. We set $b=10 \,R_J$, which corresponds to $\sim 0.1 R_H$,  within the valid regime established by \citet{Greenberg1988}.

The resulting distribution of post-encounter eccentricities is shown in Figure \ref{fig:jupiter_ejections}. The red areas correspond to areas of orbital space that would be ejected, while orbits in the blue area remain bound. All orbits with $e>0.7$ and $a<10$ au are ejected at this impact parameter, and many are scattered onto strongly hyperbolic orbits ($e'>3$). The $e'$ contours are relatively flat for low values and display increased dependence on $a$ for higher values. Note that the $\beta=1$ line exactly follows the $e'=1$ threshold represented in white.
It is worth noting that a particle with an orbit within the blue region can still be ejected. However, its orbit must evolve via multiple scatterings to be within the red region before this occurs.

In Figure \ref{fig:9panel} we show how the ejection distribution varies as a function of impact parameter over a larger range of semimajor axes. The ejection efficiency is shown for impact parameters spanning a few times the radius of the planet to over half of the Hill radius. There are several salient features of the ejection regions that are worth noting. As one would expect, closer encounters are generally more efficient. As the ejected area shrinks, $e'_{max}$ decreases rapidly; most of the $b$-values generate moderate values of $e'$. At encounter distances near 0.5 $R_H$, a particle would need a highly eccentric orbit to be ejected, and even then the orbit would be barely hyperbolic. $e'$ can be written as function of the impact parameter $b$:

\begin{equation}\label{eq:e'(b)}
\begin{aligned}
e'^2
&= 1
   - \Bigl(
        1 - U^2
        - 2U \cos\!\bigl(
            \theta
            - 2\arctan\!\left\{\tfrac{m}{bU^2}\right\}
          \bigr)
     \Bigr) \\
&\quad\;\;
   \times \Bigl(
        1 + U \cos\!\bigl(
            \theta
            - 2\arctan\!\left\{\tfrac{m}{bU^2}\right\}
          \bigr)
     \Bigr).
\end{aligned}
\end{equation}

Equation \ref{eq:e'(b)} provides a few key insights into the relationship between impact parameters and final eccentricity. $e'^2$ is at a maximum when  $\cos\!\bigl(\theta - 2\arctan\!\left\{{m/bU^2}\right\}\bigl)=1$. This is synonymous with $\gamma=\theta$ and $U'=U_y'$. In other words, the maximum eccentricity $e'_{\max}$ can be written as

\begin{equation}\label{eq:e'_max}
e'_{\max}(U) \;=\
\sqrt{\,U^3 + 3U^2 + U\,},
\end{equation}

while the associated impact parameter is given by:

\begin{equation}\label{eq:b_crit}
b_{\text{crit}} \;=\; m \bigg/\big[{U^{2} \, \tan(\theta/2)}\big].
\end{equation}

$e_{max}$ is only a function of the asymptotic speed of the particle, while $b_{crit}$ involves the mass of the planet and orientation of the interaction. $e'$ is not a monotonic function of $b$. For an orbit with $e=0.45$ and $a=a_J$, an impact parameter of $b=2 R_J$ or $b=15 R_J$ does not result in ejection, while $b=7 R_J$ does. To most effectively eject a particle via a scattering, the interaction must take place at a distance at which the final trajectory of the particle is aligned with the y-direction. In Equation \ref{eq:e_prime_eq_solved}, $e'$ does not depend on $U_x$ or $\phi$, indicating symmetry around the y-axis. $e'<e_{max}'$ for $b<b_{crit}$ because these values rotate $\vec{U}$ past the y-axis.

\begin{figure}
\centering
\includegraphics[width=1.\linewidth]{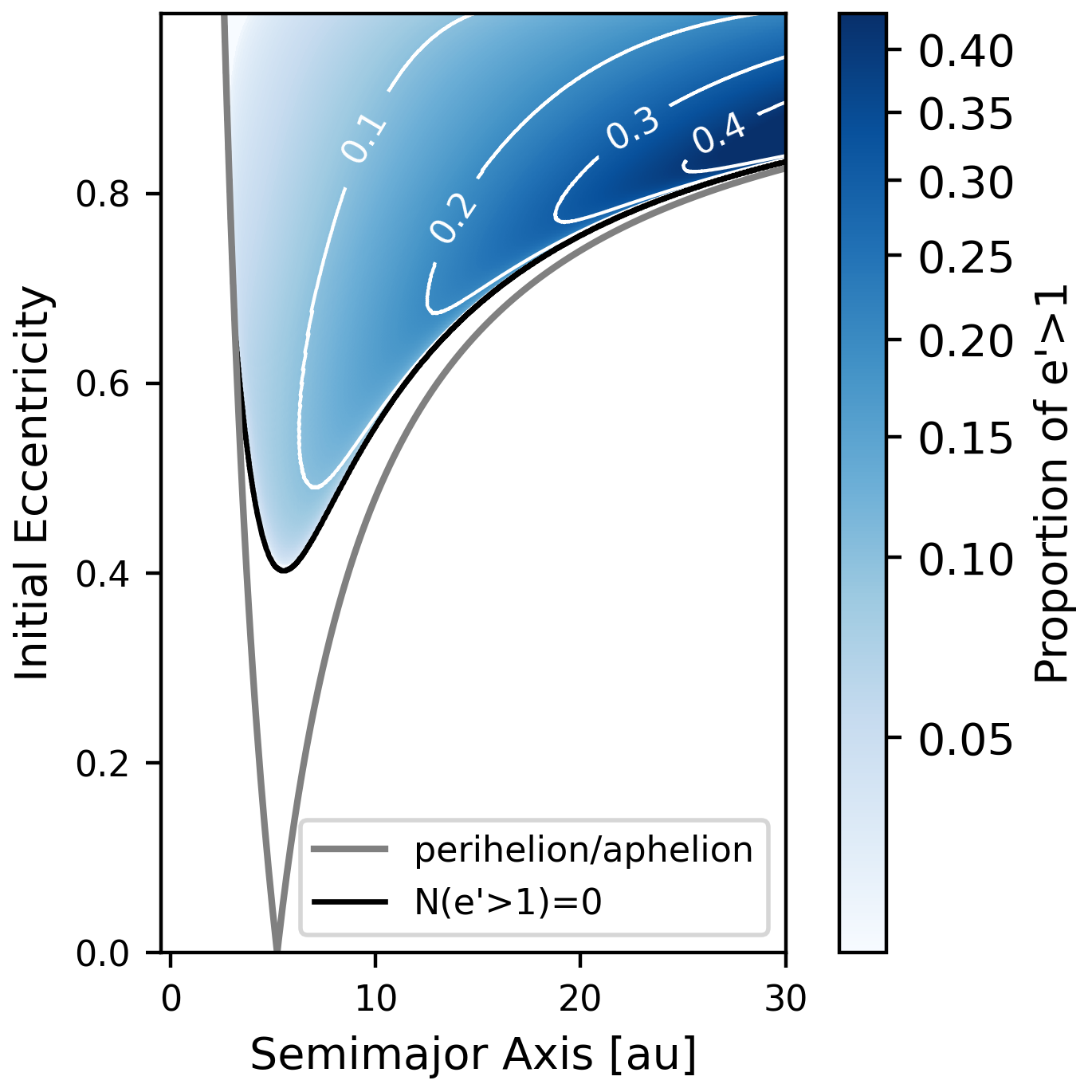}
\caption{Higher eccentricity initial trajectories are more efficiently ejected. The color indicates the total fraction of orbits that are ejected for impact parameters ranging from $R_p$ to $R_H$.}
\label{fig:impact_parameter_heatmap}
\end{figure}
\begin{figure}
\centering
\includegraphics[width=\linewidth]{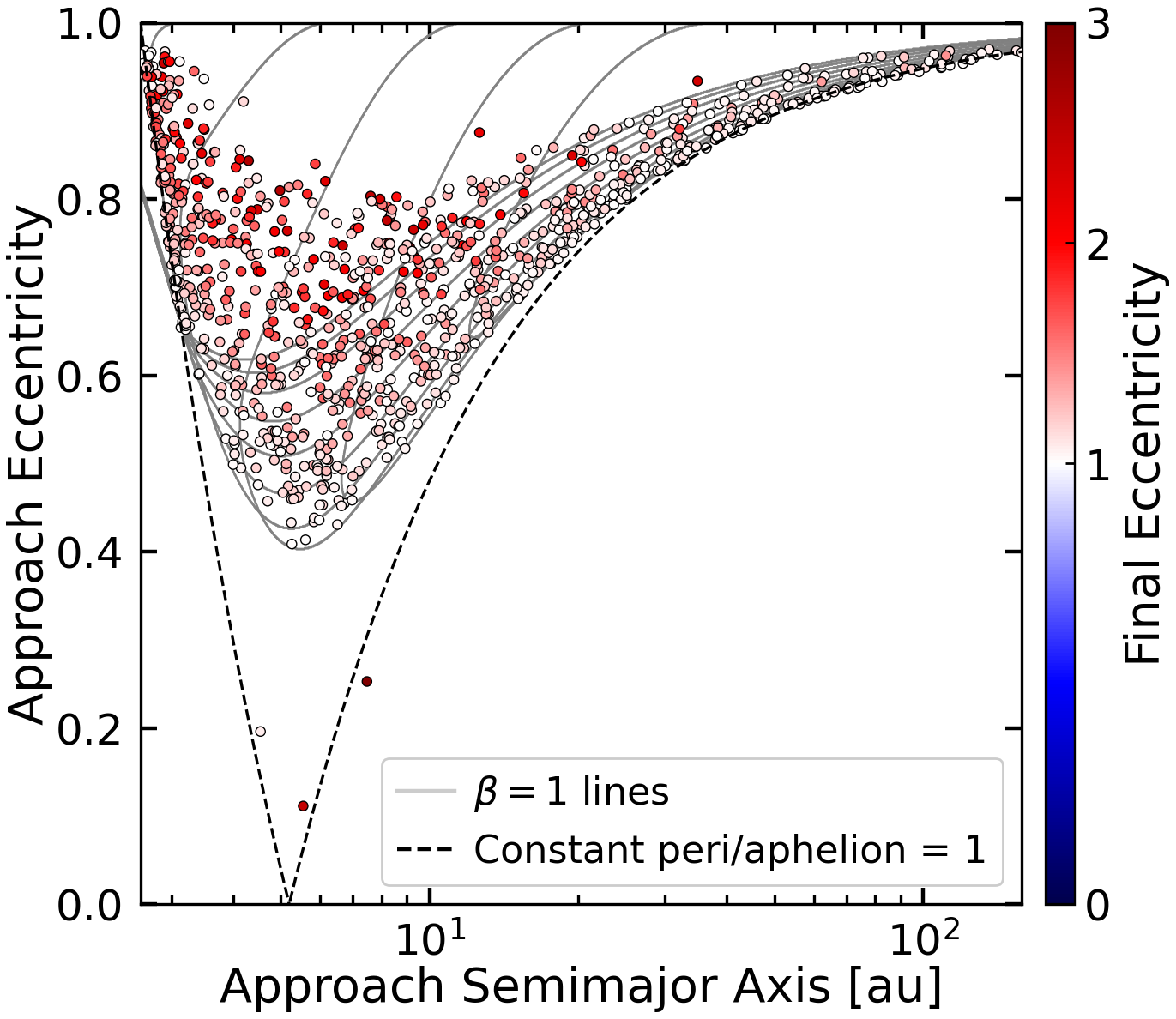}
\caption{The distribution of particles ejected from the suite of simulations of massless test particles interacting with a Jupiter analogue perturber. 1072 particles were ejected and 66 particles on highly eccentric initial orbits, $e>0.97$, were excluded. The numerical simulations provide nominal agreement with the analytic theory shown by the $\beta=1$ line. The overplotted $\beta=1$ contours are calculated from a range of $b$-values logarithmically sampled from [$R_J$,$R_H$].}
\label{fig:sim_subplot}
\end{figure}
Impact parameter can be used to characterize the distribution of particles ejected from various areas in a system. Sampling $b$-values for individual orbits gives a proportion of scatterings within a given  distance that result in ejection.  To investigate the most efficient areas of ejection in the system, we sampled $b$ linearly across [$R_J,\, R_H$]. Figure \ref{fig:impact_parameter_heatmap} shows a heatmap in $(a,\,e)$ space of the $b$-values that produce ejections. The contours broadly follow the shape of the $e'$ contours in Figure \ref{fig:jupiter_ejections}, and agree with the behavior across $b$-values shown in Figure \ref{fig:9panel}. For an orbit with $e=0.8$ and $a=30$ au, over 40\% of b-values within 1 $R_H$ are ejected. 

\begin{figure*}
\centering
\includegraphics[width=2\columnwidth]{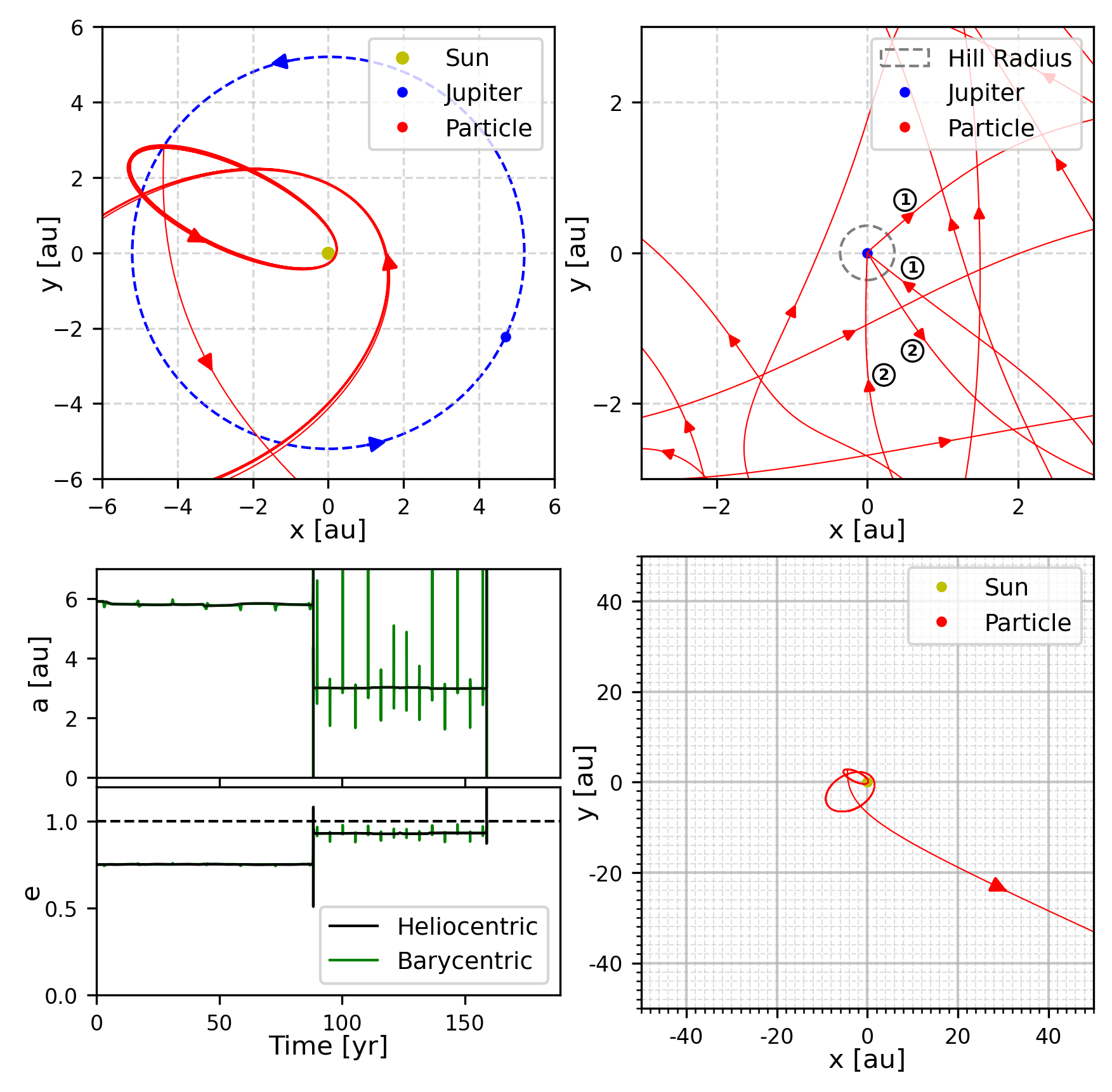}
\caption{Evolution of the orbit of a massless test particle ejected by a Jupiter analogue.  (top left) Top down view of the star, planet, and particle. The particle starts on a more distant orbit that leaves the frame of the plot, before being redirected onto a much closer orbit and is then ejected. The arrows indicate the direction of the orbits. (top right)
Top down view centered on the massive perturber. The Hill Sphere of the perturber is shown as a dotted line, and the path of the particle is shown in red. Circled numbers indicate the incoming and outgoing paths of the first and second close encounters. (bottom left) The temporal evolution of semimajor axis and eccentricity of the ejected particle in Earth years. Both heliocentric and barycentric coordinates are shown. (bottom right) Zoomed out version of the top down orbits of the star and particle. An animated version of this figure is available online.}
\label{fig:ejection_animation}
\end{figure*}

\subsection{Three-Dimensional Generalization}\label{subsec:3d}
While our two-dimensional formulation is useful in examining fundamental processes in the close-encounter ejection,  three-dimensional planetary systems are more complex. For example, the encounter probability is greatly inflated in two dimensions; two mutually inclined orbits have a lower probability of interacting. The b-plane becomes a line, and therefore the close-encounter probability is proportional to $R_H$ instead of $R_H^2$. The ejection timescales in a purely planar system, such as the one presented in Section \ref{section:simulations}, are much shorter than those of a corresponding inclined system.
However, a few key points from the two-dimensional idealized case remain applicable in the three-dimensional case.
Without assuming $U_z=0$, we use the full form of Equation \ref{eq:e_prime_eq} in this work, from Equation 32 in \citet{Carusi1990}:
\begin{equation}\label{eq:e_prime_eq_solved_3d}
 (e')^2=1-\big(1-U^2-2U_y'\big)\times\big[(U_z')^2+\big(1+U_y'\big)^2\big]\,.
\end{equation}
Notice that, because the last term is a sum of squares, Inequality \ref{eq:ej_condition} nominally remains a criterion for ejection. To show this robustly, we must check the underlying assumption laid out in Section \ref{subsec:branch_choice}. 
The derivative of Equation \ref{eq:e_prime_eq_solved_3d} w.r.t $U_y'$ (the three-dimensional version of Equation \ref{eq:e_derivative}) is:

\begin{equation}\label{e_derivative_3d}
\frac{\partial (e'^2)}{\partial U_y'} = 2\,\big(3U_y'^2+U_y'(U^2-1)+U_z^2+1\big)\,.
\end{equation}

Equation \ref{e_derivative_3d} is positive over the same interval as the planar version, so  $e'$ remains a monotonic, positively sloped function with respect to $U_y'$ over the relevant parameter space. 
In three dimensions, we have $\psi\in[0,2\pi)$ rather than $\psi\in\{0,\pi\}$ as before. Therefore, it is less obvious that $\psi=0$ generates the maximum value of $e'$. To show that this assumption holds in three dimensions, we find: 

\begin{equation}
\frac{\partial (U_y')}{\partial \psi} = -U\sin\theta\sin\gamma\sin\psi.
\end{equation}

Recall that $\sin\theta\geq0$ and $\sin\gamma\geq0$. Therefore, the sign of $\frac{\partial (U_y')}{\partial \psi}$ varies like $-\sin\psi$, giving a maximum $e'$ at $\psi=0$, as before.
Thus, the variable $\beta$ remains an analytic predictor of ejection in three dimensions. Additionally, alignment with the y-axis remains a crucial factor in ejection, although $|U_z|$ is an added consideration.

\section{Numerical Validation of Analytic Theory}\label{section:simulations}

In order to assess the applicability of our analytic conclusions, we ran a series of simulations using the REBOUND code  (\cite{Rein2012REBOUND} using the IAS15 integrator (\cite{Rein2015}). Each simulation consisted of a star with a mass of 1 $M_\odot$, one planet with {$m_P = 1 M_J$, $a_P = 1 \,a_J$, $e_P = 0$, and $i_P = 0$}, and a set of 750 massless test particles. These particles were randomly assigned a set of orbital elements with $a\in [1,10)$, $e\in[0,0.8)$, $i=0$, and $\omega,\Omega,M\in[0,2\pi)$, where $M$ is the mean anomaly. The numerical simulation  was integrated for 3000 Jovian years. While this system configuration is idealized, our goal is to present a proof-of-concept for applying the $\beta$ criterion to a specific system.

In order to avoid the error in defining the approach velocity outlined by \citet{Greenberg1988}, pre-encounter orbital elements were recorded each time the particle passed within a distance of 4 Hill Sphere radii of the planet. Upon ejection, initial elements were defined from the most recent crossing into this sphere. The results of these simulations are shown in Figure \ref{fig:sim_subplot}. The three ejected particles that had smaller initial eccentricity underwent nontrivial interactions with the planet. Specifically, these three cases were captured in the Hill sphere of the planet for many orbits.

The $\beta$ parameter can be used to evaluate orbits that can be ejected. Figure \ref{fig:sim_subplot} shows 25 $\beta=1$ contours with $b$-values sampled logarithmically from the range [$R_P$, $R_H$]. The contours outline the region of orbital element space that results in ejection including  $e\gtrsim0.4$. No single contour defines this limit across the whole range, and so it is necessary to use a set of lines to define the lower bound. Finally, objects with higher final eccentricity have higher initial eccentricity, in agreement with Figure \ref{fig:9panel}.

 Figure \ref{fig:ejection_animation} (and the corresponding animation) shows the evolution of the orbit of one of the ejected particles from these numerical experiments. As discussed in Section \ref{subsec:3d}, the ejection  timescale is much quicker than it would be for a realistic three-dimensional system. This specific particle was chosen because of a few key features of its motion. Firstly, the initial orbit of the particle is useless in characterizing the behavior during scattering, since its orbit radically shifted before the interaction. Using initial values for $a$ and $e$ rather than their approach values would have mischaracterized the interaction. Secondly, it shows the necessity of using care when using orbital elements to assess ejection of a particle. The particle in Figure \ref{fig:ejection_animation} could have been erroneously removed, since its barycentric eccentricity briefly spiked above 1 while it was still bound.

\section{Applications of methodology to Interstellar Objects}

In this section we present applications of the methodology to planetary systems. This demonstrates how the presence of multiple gravitational influences affects the ejection distribution. Characterizations of the orbits of objects with the Solar System, $\beta$ Pictoris, and HR 8799 are presented.

\subsection{Solar System Analogue}
Adding the three other Solar System giant planets to the fiducial Jupiter case allows for a similar treatment of orbits across the outer Solar System. Note that this still assumes that the massive perturbers have circular orbits. 

\begin{figure}
\centering
\includegraphics[width=1.\linewidth]{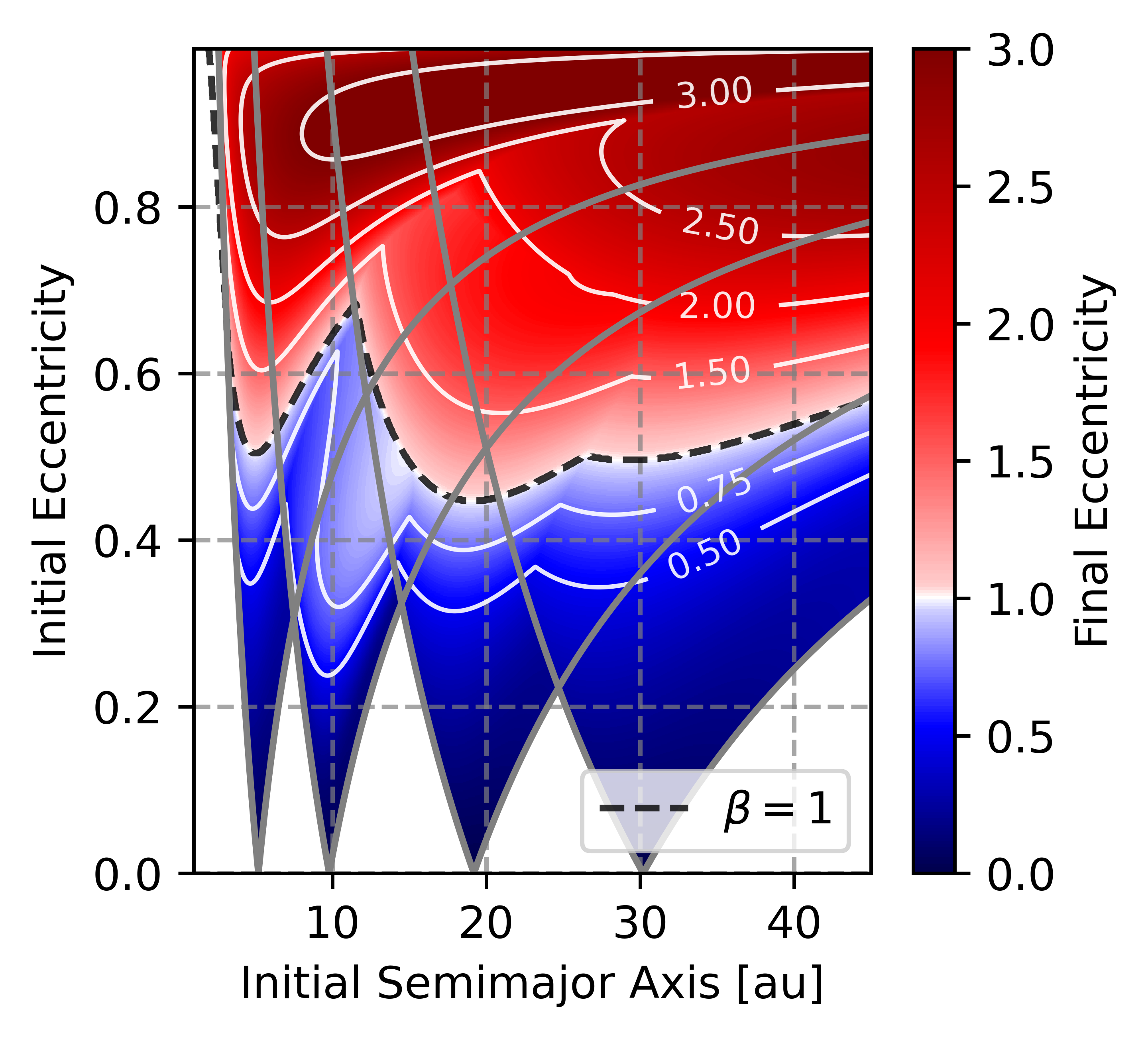}
\caption{Similar to Figure \ref{fig:jupiter_ejections} with  additional Saturn, Uranus, and Neptune analogues on circular orbits. The color indicates the maximum $e'$ for particles crossing the orbits of multiple planets. The impact parameter is set to 10 times the radius of each respective planet.}
\label{fig:SS_beta_heatmap}
\end{figure}

In Figure \ref{fig:SS_beta_heatmap} we  show the same analysis as shown in Figure \ref{fig:jupiter_ejections} but including Saturn, Uranus and Neptune analogues as well. There is a strongly positive correlation between $e$ and $e'$, and $\beta$ remains a predictor of ejection.
In Figure \ref{fig:SS_beta_heatmap}, the impact parameter is not held constant. As a direct extension of Figure \ref{fig:jupiter_ejections}, we set the impact parameter to $10 R_p$ for each planet. As such, this figure is not meant to directly compare ejection capabilities between planets.

\subsection{\texorpdfstring{$\beta$ Pictoris}{beta Pictoris}}

Two exoplanetary systems were chosen as examples of the utility of our methodology. The first system to which we applied the methodology is $\beta$ Pictoris, chosen because of its massive planets, substantial population of exocomets, and large debris disk \citep{Kiefer2014, Chilcote_2017, Feng_2022, Lagrange_2019_Pictoris}. The parent-body belt spans from around 40-150 au and has a mass of $7.5\times10^{-2}\,M_{\oplus}$, while the dust halo extends as far as 1800 au, with a mass of $1.1\times10^{-2}\,M_{\oplus}$ \citep{Ballering2011}. A constant impact parameter was used: $b=0.6\ R_{H}$ of $\beta$ Pic b, or 0.135 au.

\begin{figure}
\centering
\includegraphics[width=1.\linewidth]{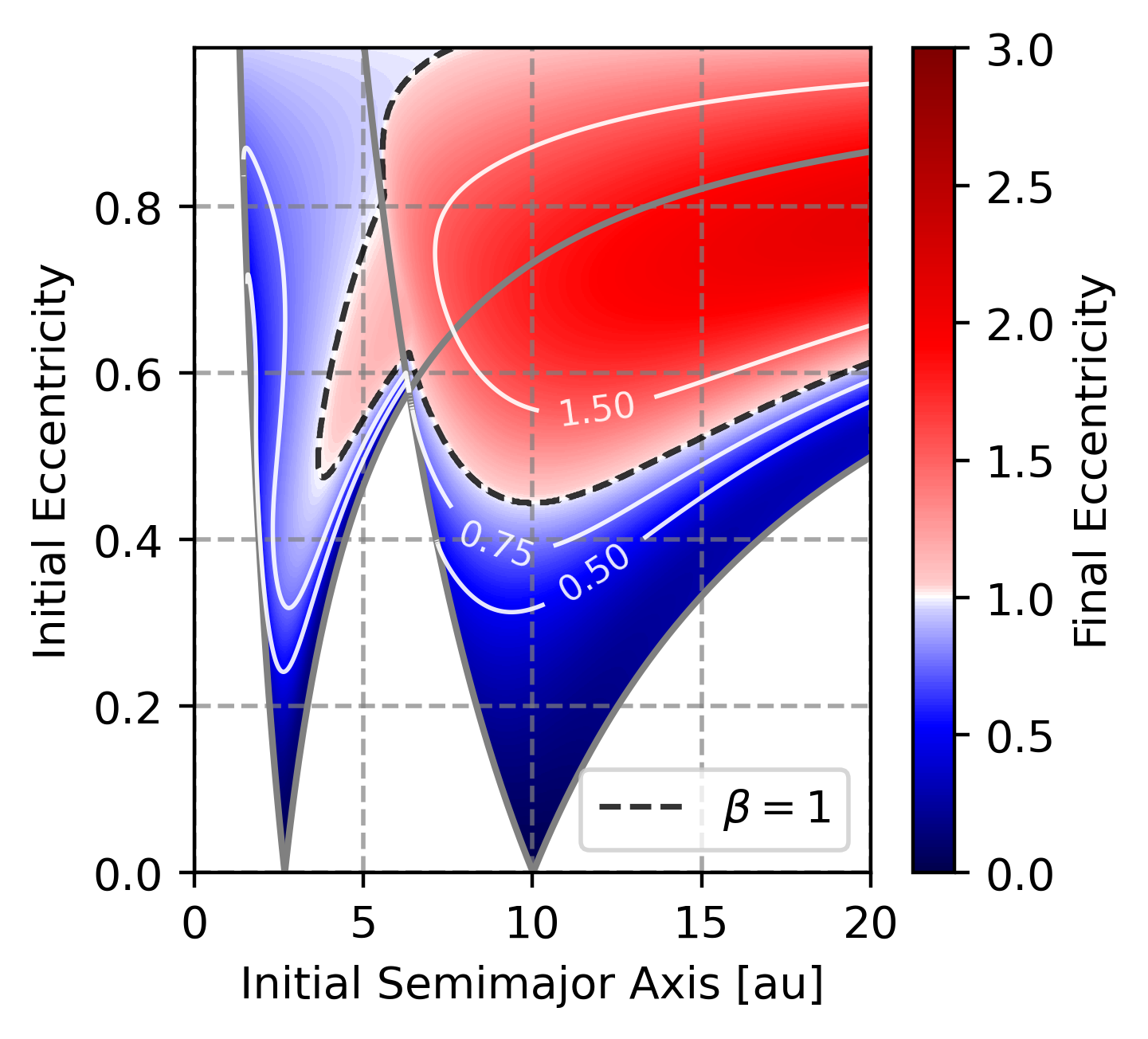}
\caption{Similar to Figure \ref{fig:jupiter_ejections} for circular versions of the planets orbiting $\beta$ Pictoris with $b$=0.135 au. The innermost planet has $a\simeq2.7$ au, $m\simeq10 M_J$, and $e\simeq0.31$, while the outermost has $m\simeq12 M_J$, $a\simeq10$ au, and $e\simeq0.11$.}
\label{fig:beta_pictoris_colormap}
\end{figure}

The differences between the ejection capability of both planets can be seen clearly in Figure \ref{fig:beta_pictoris_colormap}. Note that $\beta$ Pic b, the outermost planet, is a much better ejector than its counterpart in $\beta$ Pic c due to its greater distance from the star. Of the range of orbits with the potential to become hyperbolic at the chosen $b$-value in the $\beta$ Pictoris system, the large majority cross the orbit of $\beta$ Pic b.

\subsection{HR 8799}
The second exoplanetary system to which we applied the methodology is HR 8799. The planets in this system have low measured inclinations, high masses, and large semimajor axes \citep{Gozdziewski_2020_HR, GRAVITY_2019_HR,Marois_2010_HR, Brandt_2021_HR}. Again, a constant impact parameter was used: $b=0.1\ R_H$ of HR 8799 c, or 0.465 au. 
\begin{figure}
\centering
\includegraphics[width=1.\linewidth]{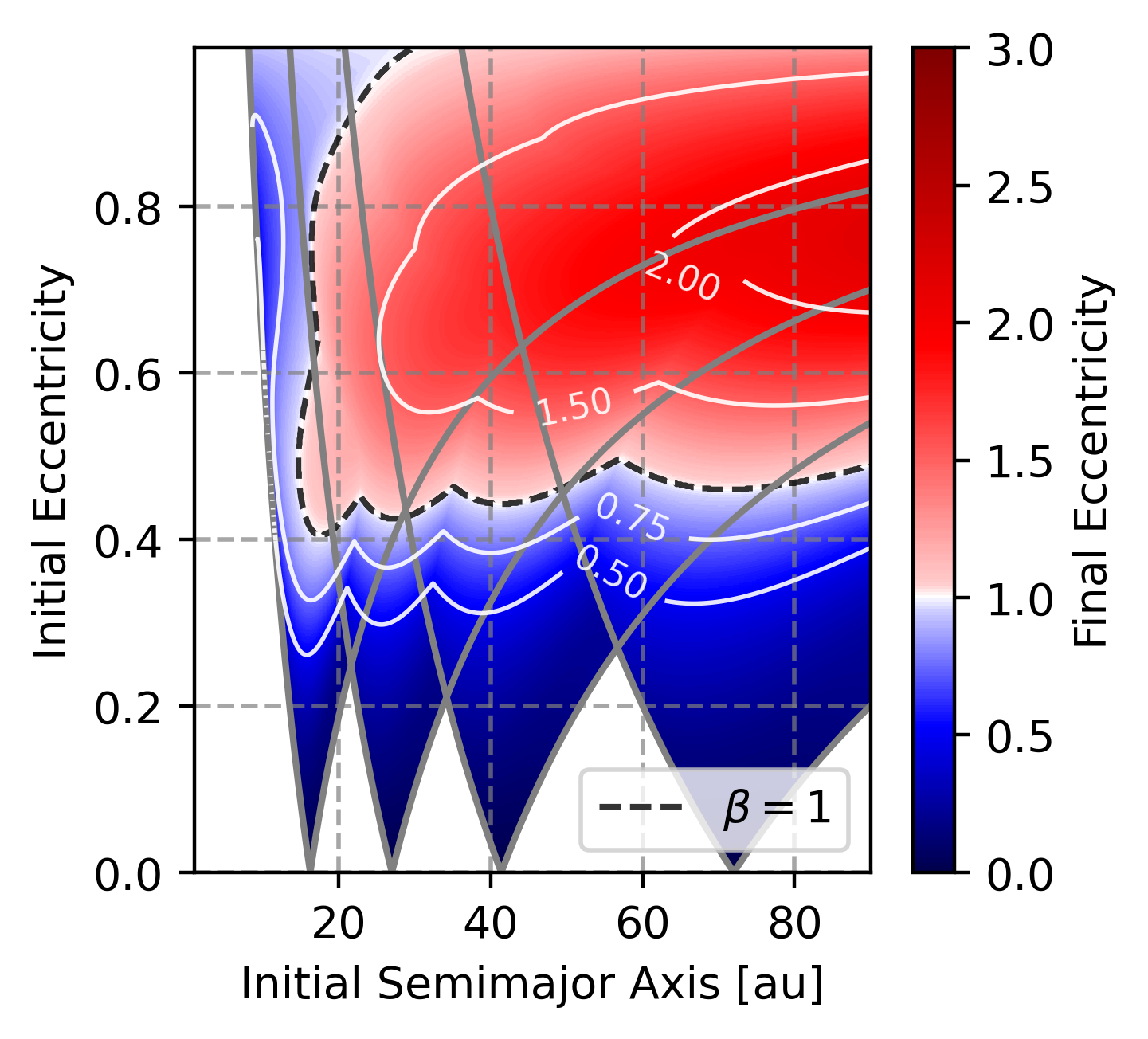}
\caption{Similar to Figure \ref{fig:jupiter_ejections} for circular versions of the planets orbiting HR 8799 with $b$=0.465 au. }
\label{fig:hr_colormap}
\end{figure}
Figure \ref{fig:hr_colormap} demonstrates  the strong influences displayed by the HR 8799 planets. An impact parameter more than three times greater than the one used for $\beta$ Pictoris still generates greater final eccentricities. This is largely due to the much greater semimajor axes of the HR 8799 planets. The influences of the planet are of similar strengths, unlike the $\beta$ Pictoris planets. Most particles with initial eccentricity larger than $e\sim0.4$ will be ejected from the regions of HR8799 that host planets.

\subsection{Numerical Simulations of HR 8799}
We perform an additional set of numerical simulations to validate the exoplanet case of HR 8799. The planets were instantiated with orbital elements taken from Table 7 and Model 1 of \citet{Zurlo_2022_HR} to preserve stability to $t=\infty$. Test particles were initialized with $a\in [15,75)$ au, $e\in[0,0.8)$, $i=26.8734271$, $\Omega=62.1852189$, and $\omega,M\in[0,2\pi)$. 

\begin{figure}
\centering
\includegraphics[width=1.\linewidth]{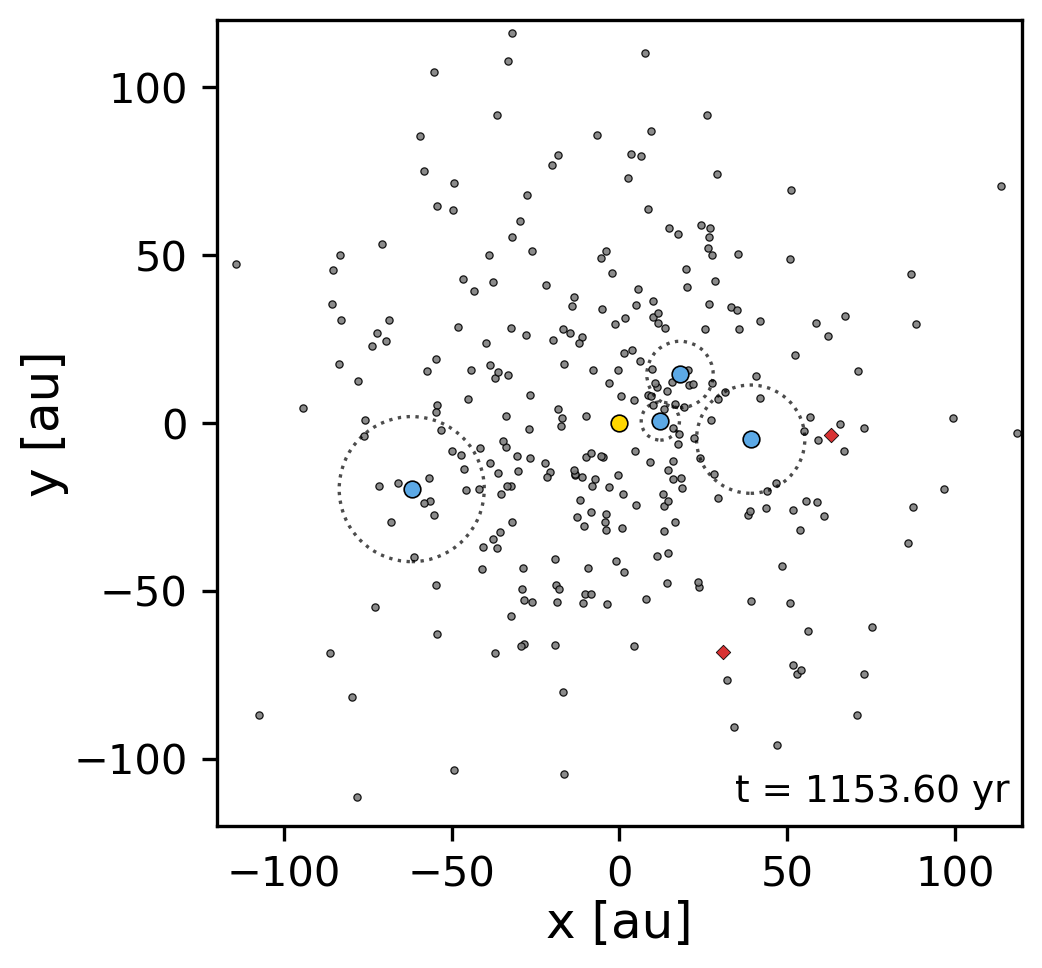}
\caption{A frame of the HR 8799 simulation. The blue circles represent the planets, with their corresponding 3 $R_H$ sphere shown as a dotted line. This frame is taken from early in the simulation and shows two particles in the process of being ejected, denoted by the red diamonds. An animated version of this figure showing the first $\sim$1000 Earth years of the simulation is available online.}
\label{fig:hr_sim}
\end{figure}

\begin{figure*}
\centering
\includegraphics[width=1.\linewidth]{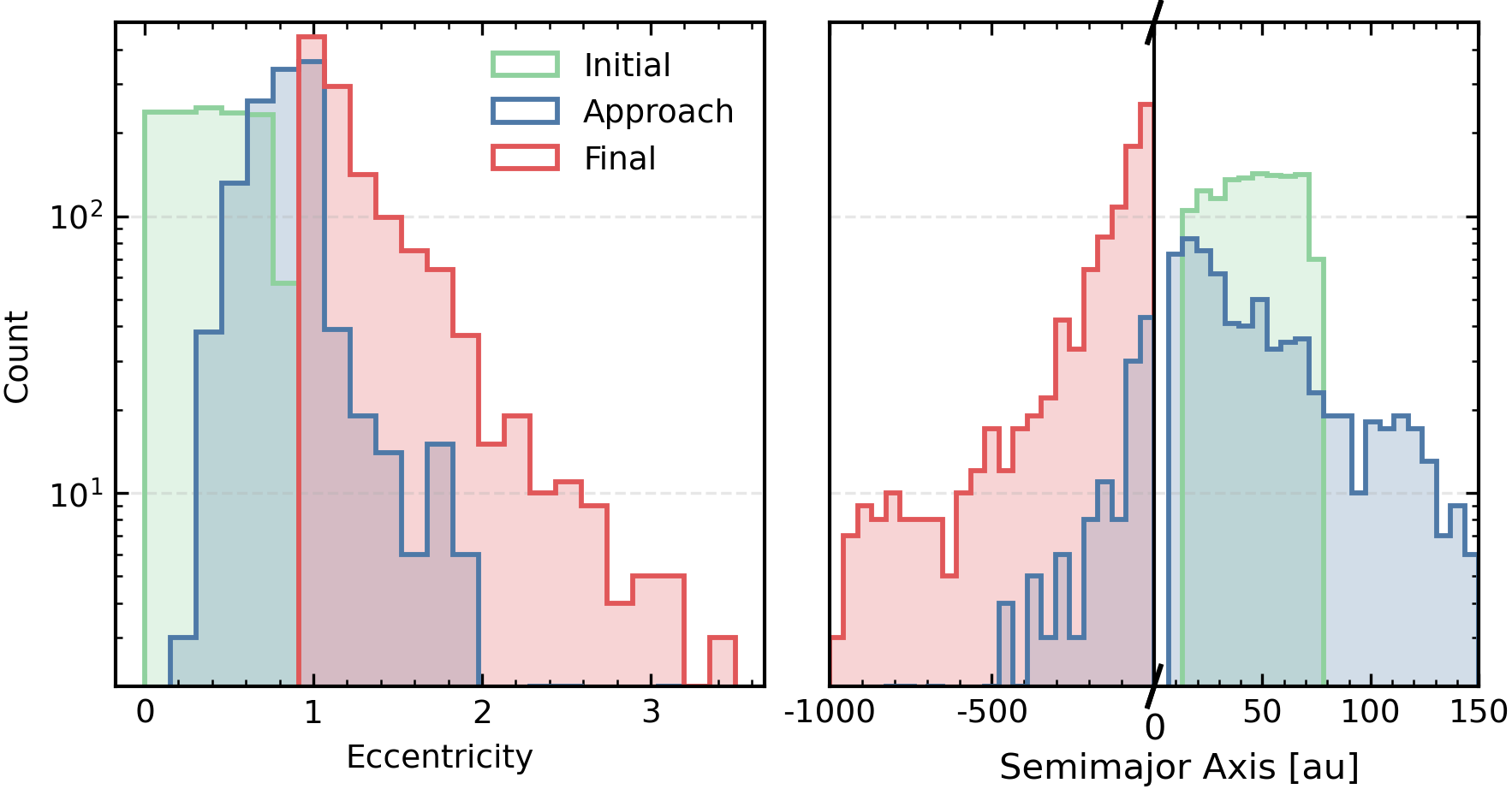}
\caption{The distributions of eccentricity and semimajor axis of test particles in the simulated HR 8799 system. 1248 of 1250 particles were ejected over $\sim$135,000 Earth years. }
\label{fig:hr_hist}
\end{figure*}

The simulation is largely the same as the single-planet instance described in Section \ref{section:simulations}. For each planet, the zone of influence was defined as $3 R_{HS}$. Figure \ref{fig:hr_sim} shows a snapshot of the simulation, and an accompanying animation of the entire simulation is available online. Figure \ref{fig:hr_hist} shows the initial, approach, and final semimajor axis and eccentricity of the particles in the simulation.

 Almost every single test particle that started bound to the system is ejected after the course of the simulation (Figure \ref{fig:hr_hist}). The distribution of semimajor axis broadens significantly. Moreover, a fraction of the particles already have $e>1$ when they approach a planet. This suggests that they were scattered by one planet before quickly entering the influence of another. This kind of quick exchange does not get flagged as an ejection until the particle completely leaves the influence of all the planets. This demonstrates that the effects of each planet are compounded to create more complex interactions. There is a significant shift between the initial distribution and the approach distribution, which indicates that most particles are scattered, perhaps multiple times, before undergoing the final interaction that ejects them.

\section{Discussion}
In this paper we calculate pre- and post-encounter orbital elements of particles ejected from planetary systems in simple, representative cases. Building on the methodology presented by \citet{Opik1976}, we have presented a mathematical framework that maps pre-encounter orbital elements to post-encounter outcomes. This methodology may provide insight into the process of ejection by close-encounter scattering, even in more complicated cases.

 Particle ejection is closely tied with the y-component of the planet-centric velocity. Because the y-axis is aligned with the velocity vector of the planet, an increase in the y-direction velocity of a particle represents an increase in orbital energy in the star-centric frame.

A key assumption in our calculations is that the encounter occurs in the ecliptic plane. In other words, this method assumes that the difference between the inclinations of the perturber and particle is negligible. This simplification is restrictive on an encounter-by-encounter basis. Therefore, the methodology presented here is meant to generate broad estimates of ejection efficiency and pre-ejection orbital characteristics. Extending this framework to include the full three-dimensional scattering geometry, where inclination and argument of pericenter are free parameters, would allow for a more complete characterization of the phase space of ejection. Crucially, a significant portion of our findings - namely the definition of $\beta$ for use as an ejection marker, and our designation of $\psi=0$ and alignment with the y-axis as most favorable to ejection - hold in the three-dimensional case as well.

This methodology can be applied to several different scenarios. It offers an efficient mechanism to compute post-encounter trajectories without employing direct numerical integrations. This makes it a useful tool to rapidly evaluate the efficacy with which a given exoplanetary system will eject comets, and where within that system those comets are most likely to be ejected from. The method could be applied to star-star and star-black hole interactions. 

Our methodology is related to that of \citet{Huang2025} who presented a complementary analysis regarding particle ejection.  Those authors focused on random walks of the orbital energy of planetesimals, and calculated the scattering timescale, dynamical lifetime, and ejection speed. Their work characterizes the evolution of small-body orbits that leads up to the final scattering event. See Section 6 in that work for discussion about applications to interstellar objects.

It has been demonstrated that circumbinary systems (and especially misaligned circumbinaries) are especially efficient progenitors of interstellar objects (\cite{Cuk2017,Jackson2017,Childs2022}). This methodology could be directly applicable to investigating ejection of interstellar objects from circumbinaries.

\section*{Acknowledgments}

We thank the anonymous reviewer for an insightful referee report that strengthened the scientific content of this manuscript.

We thank Mackenzie Ticoras, Brian O'Shea,  Dong Lai, and Lieke van Son for useful conversations. 

H. M. acknowledges support from the Lawrence W. Hantel Endowment in Memory of Prof. Donald J. Montgomery.

D.Z.S. is supported by an NSF Astronomy and Astrophysics Postdoctoral Fellowship under award AST-2303553. This research award is partially funded by a generous gift of Charles Simonyi to the NSF Division of Astronomical Sciences. The award is made in recognition of significant contributions to Rubin Observatory’s Legacy Survey of Space and Time.

\section*{Data Availability}

All  code used for generating the simulations in this paper are available at \url{https://github.com/haydenmonk/Hyperbolic_trajectories_code}.

\bibliographystyle{mnras}
\bibliography{main}

\bsp	
\label{lastpage}
\end{document}